\documentclass[onecolumn,prd,aps,nofootinbib]{revtex4}
\usepackage{graphicx}
\usepackage{dcolumn}
\usepackage{bm}
\usepackage{amsmath}
\usepackage{epsfig}
\usepackage{graphicx}
\usepackage{hyperref}
\usepackage{color}
\usepackage{url}
\usepackage{multirow}

\def\be{\begin{equation}}
\def\ee{\end{equation}}
\def\bea{\begin{eqnarray}}
\def\eea{\end{eqnarray}}

\usepackage{color}

\begin{document}

\title{Cold Dark Matter Isocurvatures Perturbations: Cosmological Constraints and Applications}

\author{Hong Li$^{a,b}$}
\email{hongli@ihep.ac.cn}
\author{Jie Liu${^a}$}
\email{liujie@ihep.ac.cn}
\author{Jun-Qing Xia${^c}$}
\author{Yi-Fu Cai${^d}$}

\affiliation{${}^a$Institute of High Energy Physics, Chinese Academy
of Science, P.O. Box 918-4, Beijing 100049, P. R. China}

\affiliation{${}^b$Theoretical Physics Center for Science Facilities
(TPCSF), Chinese Academy of Science, Beijing 100049, P.R.China}

\affiliation{${}^c$Scuola Internazionale Superiore di Studi
Avanzati, Via Bonomea 265, I-34136 Trieste, Italy}

\affiliation{${}^d$Department of Physics, Arizona State University,
Tempe, AZ 85287, USA}

\begin{abstract}

In this paper we present the constraints on cold dark matter (CDM) isocurvature contributions to the cosmological perturbations. By employing Markov Chain Monte Carlo method (MCMC), we perform a global analysis for cosmological parameters using the latest astronomical data, such as 7-year Wilkinson Microwave Anisotropy Probe (WMAP7) observations, matter power spectrum from the Sloan Digital Sky Survey (SDSS) luminous red galaxies (LRG), and ``Union2'' type Ia Supernovae (SNIa) sample. We find that the correlated mixture of adiabatic and isocurvature modes are mildly better fitting to the current data than the pure adiabatic ones, with the minimal $\chi^2$ given by the likelihood analysis being reduced by $3.5$. We also obtain a tight limit on the fraction of the CDM isocurvature contributions, which should be less than $14.6\%$ at $95\%$ confidence level. With the presence of the isocurvature modes, the adiabatic spectral index becomes slightly bigger, $n_s^{\rm adi}=0.972\pm0.014~(1\,\sigma)$, and the tilt for isocurvature spectrum could be large, namely, the best fit value is $n_s^{\rm iso}=3.020$. Finally, we discuss the effect on WMAP normalization priors, shift parameter $R$, acoustic scale $l_A$ and $z_{*}$, from the CDM isocurvaure perturbation. By fitting the mixed initial condition to the combined data, we find the mean values of $R$, $l_A$ and $z_{*}$ can be changed about $2.9\sigma$, $2.8\sigma$ and $1.5\sigma$ respectively, comparing with those obtained in the pure adiabatic condition.

\end{abstract}

\maketitle

\section{introduction}
\label{intro}

The accumulation of WMAP seven year measurement on cosmic microwave background radiation (CMB) \cite{wmap7}, associated with observations of SDSS \cite{sdsslrg7}, provide wealthy information on the anisotropies and inhomogeneities of our universe, in light of which the perturbation theory has been tested in certain level. Currently, various observational data favor the simplest concordance cosmological model, which has six free parameters and the pure adiabatic initial condition \cite{wmap7,Xia:2008ex,Li:2006ev,Li:2008vf,Li:2010ac}. Although the concordance model fits the data quite well, it is always worthy to consider alternative candidates. Namely, it is important to study observational constraints on initial states of cosmological perturbations at reheating surface.

Generically, there exists two classes of modes of cosmological perturbations, with one being adiabatic of which the trajectory is parallel to the background evolution, while the other isocurvature of which the trajectory is orthogonal to the background evolution\cite{Linde:1985yf, Kofman:1986wm, Mollerach:1990ue, Kawasaki:1995ta, Polarski:1994rz, Sasaki:1995aw, GarciaBellido:1995qq, Gordon:2000hv, Bartolo:2001rt}. However, a first lesson from observational data is that the primordial fluctuations are nearly adiabatic; additionally, an isocurvature mode is also expected to be negligible as is predicted by the simplest inflation model in terms of a single inflaton field and a rapidly reheating process\cite{Kofman:1997yn}. Therefore, one usually makes the data fitting without considering the isocurvature mode.

Accompanied with developments of inflationary cosmology, models of multiple field inflation were extensively studied in the literature\cite{Liddle:1998jc, Kanti:1999vt, Copeland:1999cs, Langlois:1999dw, Wands:2002bn, Piao:2002vf, Dimopoulos:2005ac, Langlois:2008wt, Cai:2009hw, Cai:2008if, Cai:2010wt}, which predicted an existence of primordial isocurvature fluctuations. These primordial isocurvature modes could be transferred into cosmological perturbations after reheating, such as Bayon, CDM, DE, and neutrino respectively\cite{Seckel:1985tj, Linde:1991km, Linde:1996gt,Liu:2010ba}. Consequently, the attention on isocurvature modes has been awaken in recent years\cite{Bucher:1999re, Challinor:1998xk}.

One may notice that, although the pure isocurvature primordial perturbation has been
ruled out by the Boomerang and MAXIMA-1 data \cite{Enqvist:2000hp} already, a mixture of adiabatic and isocurvature modes can be in agreement with the current data fortunately. In the literature \cite{Gordon:2002gv, Crotty:2003rz, Bucher:2004an,
Moodley:2004nz, KurkiSuonio:2004mn, Beltran:2005xd, Bean:2006qz, Trotta:2006ww, Sollom:2009vd, Valiviita:2009bp,Beltran:2005gr,Mangilli:2010ut}, the studies on constraining the isocurvature fluctuation have been performed extensively using various observational data, such as CMB, LSS, integrated Sachs-Wolfe effect or Lyman-$\alpha$ forest data. With the WMAP7 data, the totally un-correlated and anti-correlated adiabatic, non-adiabatic perturbations are constrained \cite{wmap7}, which are performed by fixing the correlation coefficients to be $0$ or $-1$, respectively.

In this paper, we study the constraints on the isocurvature modes of cosmological perturbations in light of the latest observational data. We consider a generic scenario that the adiabatic modes and isocurvature ones are allowed mixed through a correlation matrix which could be arising from a time-varying background trajectory. In the detailed analysis, we treat the coefficients of correlation matrix as free parameter and make the data fitting. Comparing with the previous results in the literature, our result shows a slight improvement on the final constraints of the mixed initial condition parameters. Namely, CDM isocurvature components are stringently limited, the contribution from the isocurvature modes are only allowed in small scales, also the error bars of initial condition parameters become smaller than the past works in the literature, which can be observed from the one dimensional probability distribution of the fraction parameter $\alpha$ as analyzed in the main context.

The WAMP normalization priors, $R$, $l_A$ and $z_{*}$ encoding the information of background cosmic distances, can be applied to greatly simplify the numerical calculations of determining cosmological parameters, such as the EoS of dark energy. It was found that these priors could be sensitive to the peak locations and local structures of the CMB temperature power spectrum \cite{Li:2008cj}. However, it is well-known that these quantities can be affected by CDM isocurvature perturbation. Therefore, we study the effects on the WMAP normalization priors given by the WMAP group from the isocurvature mode perturbation.

The outline of this paper is as follows. In Section II, we describe the parametrization of cosmological perturbations with adiabatic and isocurvature modes mixed, and then we consider a specific example to illustrate the effects of isocurvature perturbation imprinted on the CMB temperature power spectrum and LSS matter power spectrum respectively. In Section III, we perform a global analysis and introduce the data we applied and the parameters used in the data fitting. The constraints on these parameters are present in detail in Section IV, and the corresponding effects on reduced distance parameters are discussed. Finally, Section V includes the conclusions and discussion.

\section{Parametrization of Correlated Adiabatic and Isocurvature Perturbations}

For model-independent consideration, the initial conditions for
the correlated mixture of adiabatic and isocurvature modes can be
written as \be\label{primordial_pk}\mathcal{P}^{\rm ij}=A^{\rm
ij}_s\left(\frac{k}{k_0}\right)^{n^{\rm ij}_s-1}~,\ee where $k_0$ is
the pivot scale. Here, $A_s^{\rm ij}$ and $n^{\rm ij}_s$ are
$2\times2$ symmetric matrices which characterize the amplitude and
power spectrum index, respectively. We have
 \bea A_s^{\rm ij}=\begin{pmatrix}
A_s^{\rm adi}&\sqrt{A_s^{\rm adi}A_s^{\rm iso}}\cos\Delta\\
 \sqrt{A_s^{\rm adi}A_s^{\rm iso}}\cos\Delta & A_s^{\rm iso}
\end{pmatrix}~,
 \eea where $\cos\Delta={A_s^{\rm adi,iso}}/{\sqrt{A_s^{\rm
adi}A_s^{\rm iso}}}$ describes the correlation between the adiabatic
mode and the isocurvature mode \cite{Langlois:1999dw}, $A_s^{\rm
adi}$ and $A_s^{\rm iso}$ are the amplitude of adiabatic and
isocurvature modes, respectively. The index matrix $n^{ij}_s$ is
given by: \bea n_s^{\rm ij}=\begin{pmatrix}
n_s^{\rm adi}&n^{\rm cor}_s\\
 n^{\rm cor}_s & n_s^{\rm iso}
\end{pmatrix}~,
 \eea
where $n^{\rm adi}_s$ and $n^{\rm iso}_s$ are the spectrum indices
for adiabatic and isocurvaure modes. Following Ref.
\cite{KurkiSuonio:2004mn}, we assume $n^{\rm cor}_s={(n_s^{\rm
adi}+n_s^{\rm iso})}/{2}$ for simplicity.

Both adiabatic and isocurvature perturbations seed the structure
formation of our universe, and can be imprinted in CMB temperature,
polarization power spectrum, as well as matter power spectrum of
galaxies surveys. Symbolically, the anisotropies of CMB photons can
be written as \be\frac{\delta T}{T}= \left(\frac{\delta
T}{T}\right)_{\rm adi}+\left(\frac{\delta T}{T}\right)_{\rm
iso}+\left(\frac{\delta T}{T}\right)_{\rm corr}~.\ee Consequently,
the CMB power spectrum is given by
\begin{eqnarray}
{C}_\ell^{\rm TT} = A_s^{\rm
adi}\hat{C}_\ell^{\rm adi}+A_s^{\rm iso}\hat{C}_\ell^{\rm
iso}
+2\sqrt{A_s^{\rm adi}A_s^{\rm iso}}\hat{C}_\ell^{\rm
adi,iso}\cos\Delta~,
\end{eqnarray}
where initial power spectrum can be obtained as follows, \be\hat{C}_\ell^{\rm
ij}=\frac{4\pi}{2\ell+1}\int d\ln
k\left(\frac{k}{k_0}\right)^{n_s^{\rm ij}-1}\Theta_\ell^{\rm
i}(k)\Theta_\ell^{\rm j}(k)~,\ee with $i$, $j$ standing for adiabatic or isocurvature and $\Theta_\ell^{\rm x}$ are the
transfer functions of the radiation at initial moment. There are
similar formulas for CMB polarization power spectra $C^{\rm
EE}_\ell$, $C^{\rm BB}_\ell$ and temperature-polarization cross
correlation power spectrum $C^{\rm TE}_\ell$.

Moreover, the matter power spectrum $P(k)$ can be written as:
\begin{eqnarray}\label{pk}
P(k) = A_s^{\rm adi}\hat{P}^{\rm adi}(k)+A_s^{\rm iso}\hat{P}^{\rm iso}(k) +2\sqrt{A_s^{\rm adi}A_s^{\rm iso}}\hat{P}^{\rm adi,iso}(k)\cos\Delta ~,
\end{eqnarray}
with \be \hat{P}^{\rm ij}(k) =
\left(\frac{k}{k_0}\right)^{n_s^{\rm ij}-1}T^{\rm i}(k)T^{\rm
j}(k)~, \ee where $T^{\rm x}(k)$ are the transfer functions of
matter component at initial moment.

\begin{figure}[htbp]
\centering
\includegraphics[width=0.7\linewidth]{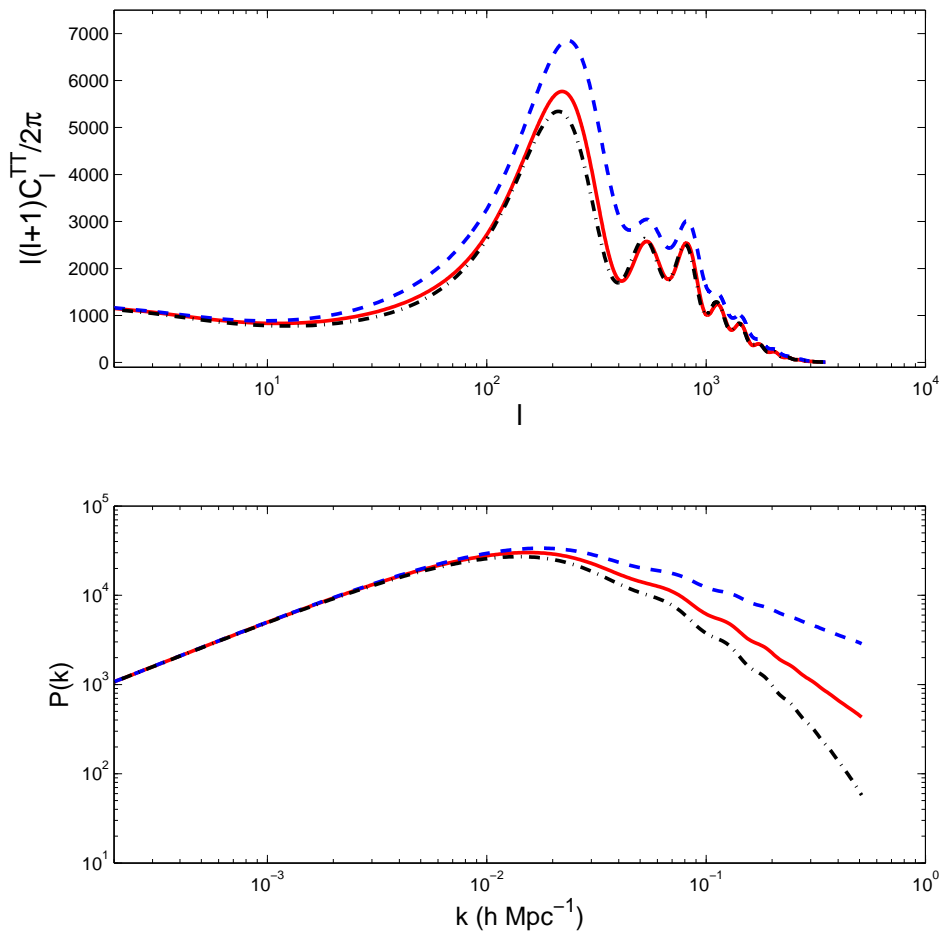}
\caption{The CMB temperature power spectra and LSS matter power
spectra are plotted using different initial conditions in the upper
and lower panels, respectively. The red solid lines are given by the
best fit model with the mixed initial condition. The blue dash lines
and the black dash-dot lines are given by choosing $A_s^{\rm
iso}=1\times10^{-10}$ and $\cos\Delta=-1$, respectively.}
\label{inprinting_cmb_pk}
\end{figure}

In order to show the effect of isocurvature perturbations, in
Fig.\ref{inprinting_cmb_pk} we plot the CMB temperature power
spectra (upper panels) and LSS matter power spectra (lower panel).
The red solid lines are given by the best fit model with mixed
initial conditions by fitting the current observational data:
$\omega_b=0.0226$, $\omega_c=0.1144$, $A_s^{\rm
adi}=2.407\times10^{-9}$, $A_s^{\rm iso}=3.885\times10^{-12}$,
$n_s^{\rm adi}=0.968$, $n_s^{\rm iso}=3.020$, $\cos\Delta=0.149$.
The blue dash line and black dash-dot lines are given by choosing
$A_s^{\rm iso}=1\times10^{-10}$ and $\cos\Delta=-1$ instead,
respectively. One can see that by taking into account the CDM
isocurvaure modes and its correlation with adiabatic perturbations, the
hight and location of the peaks of TT spectra will be modified,
and the amplitude of matter power spectrum on small scales are
changed significantly.

\section{Observational constraints}
\label{approach}

\subsection{Data}

We perform a global analysis by employing a modified MCMC package CosmoMC \footnote{\url{http://cosmologist.info/cosmomc/.}} \cite{cosmomc}, in which we have made an extension of the initial perturbations by including correlated adiabatic and CDM isocurvature modes. In computation of CMB we have included the WMAP7 temperature
and polarization power spectra with the routine for computing the likelihood supplied by the WMAP team\footnote{\url{http://lambda.gsfc.nasa.gov/.}} Since the CDM isocurvature perturbations affect the power spectrum on small scales, we also include some small-scale temperature anisotropies measurements, ACBAR \cite{acbar}, CBI \cite{cbi} and Boomerang \cite{boomerang}. Furthermore, we use the matter power spectrum measured by the observations of luminous red galaxie (LRG) from SDSS data release seven \cite{sdsslrg7}, and the ``Union II'' supernovae datasets calibrated by SALT2 template \cite{union2}.

\subsection{Parameters}

As pointed out in Refs.\cite{KurkiSuonio:2004mn, Valiviita:2009bp},
the constraints of the spectral index may depend on the pivot scale.
In order to avoid such dependence, we do not take the spectra
indices $n_s^{\rm ij}$ as free parameters in our analysis. We take
two different amplitudes $A_s(k_1)$ and $A_s(k_2)$ at two pivot
scales $k_1=0.002$ and $k_2=0.05$ to be free. And the spectra
indices can be derived by \be n_s^{\rm ij}-1 \equiv
\frac{\ln[\mathcal{P}^{\rm ij}(k_2)/\mathcal{P}^{\rm
ij}(k_1)]}{\ln[k_2/k_1]} =
\frac{\ln[A_s^{ij}(k_2)/A_s^{ij}(k_1)]}{\ln[k_2/k_1]}~. \label{ns}\ee We
assume that the overall amplitude is composed of two components: \be
A_s\equiv A_s^{\rm adi}+A_s^{\rm iso}~, \label{AS}\ee and the fraction of
adiabatic part is defined by \be\label{alpha}\alpha\equiv{A_s^{\rm adi}}/{A_s}.\ee
Thus, our most general parameter space is \be \mathcal{\bf
P}\equiv\left\{\omega_b,\omega_c,\Theta_s,\tau, \cos\Delta,
\alpha_1, \alpha_2, A_s(k_1), A_s(k_2)\right\}. \ee where
$\omega_b\equiv\Omega_bh^2$, $\omega_c\equiv\Omega_bh^2$ denotes
physical baryon density and cold dark matter density relative to
critical density respectively; $\Theta_s\equiv100\frac{r_s}{d_A}$ is
the ratio of sound horizon to angular diameter distance at
decoupling, $\tau$ characterizes the optical depth to reionization.
$A_s(k_1)$, $A_s(k_2)$, $\alpha_1$ and $\alpha_2$ are the amplitudes
and the fractions of adiabatic mode at two scales, respectively. In
our calculation, we assume the flat universe and dark energy is the
cosmological constant.

\begin{table}[htbp]
\centering\caption{The mean values and $1~\sigma$ errors of $A_s$ given by EQ. \ref{AS} at two different pivot scale, $\alpha_2$ and $\cos\Delta$, while the limit of $\alpha_1$ is for $95\%$ confidence level.}
\begin{tabular}{c|c|c|c|c}
\hline\hline $\log[10^{10}A_s(k_1)]$ & $\log[10^{10}A_s(k_2)]$ & $\alpha_1$ & $\alpha_2$ & $\cos\Delta$ \\
\hline
 $3.185_{-0.042}^{+0.041}$ & $3.622_{-0.247}^{+0.242}$ & $>93.9\%$ & $0.594_{-0.083}^{+0.056}$& $0.094^{+0.075}_{-0.095}$\\
\hline
\end{tabular}

\end{table}

\begin{widetext}
\begin{center}
\begin{table}[htbp]
\centering\caption{The mean values and $1\,\sigma$ errors of the
parameters related to the mixed initial adiabatic and CDM
isocurvature perturbations. \label{table_cp}}
\begin{tabular}{c|c|c|c|c|c|c}
\hline\hline & $\log[10^{10}A_s^{\rm adi}]$ & $n_s^{\rm adi}$ &
$\log[10^{10}A_s^{\rm iso}]$
& $n_s^{\rm iso}$ & $\cos\Delta$ & $\chi^2_{\rm min}$ \\
\hline
Adiabatic& $3.204\pm0.036$ & $0.964\pm0.010$ & -- & --& -- & $8217.0$\\
\hline Mixed & $3.165\pm0.044$ & $0.972\pm0.014$ &
$-1.375^{+1.233}_{-1.306}$ &
$2.246^{+0.494}_{-0.428}$ & $0.094^{+0.075}_{-0.095}$ &$8213.5$\\
\hline
\end{tabular}\end{table}
\end{center}
\end{widetext}

\begin{figure}[htbp]
\includegraphics[width=0.9\linewidth]{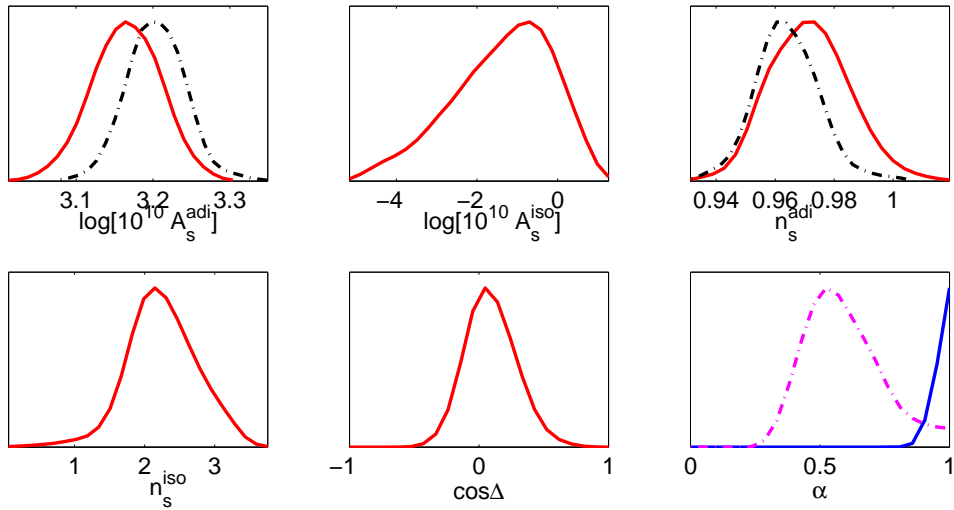}
\caption{One dimensional probability distributions of initial
conditions associated parameters are plotted with red solid lines.
For $\log[10^{10} A_s^{\rm adi}]$ and $n_s^{\rm adi}$, we specially
make a comparison with the results obtained from the pure adiabatic
initial conditions which are plotted with black dash lines. We also
plot the adiabatic perturbation modes fraction parameter $\alpha_1$
and $\alpha_2$ using the purple dash-dot line and blue solid line,
respectively.}\label{fig_isopar}
\end{figure}

\section{Global Fitting Results}

In this section, we present the main constraints on the cosmological
parameters, as well as the parameters related to the CDM
isocurvature perturbations. Then, we discuss the effects on
the WMAP normalization priors when taking into account
CDM isocurvature perturbations.

\subsection{Constraints on cosmological parameters}
\label{results_cp}

In table I and table\ref{table_cp} we list the constraints on the initial condition associated parameters by fitting the latest observational data. The parameters presented in table I are the free parameters during our calculation, while those in table \ref{table_cp} are the derived parameters from the those in table I according to EQ. (\ref{ns}-\ref{alpha}), and both results are given by the statistics of the samples given by our MCMC approach. From the mixed initial conditions, the constraints on
the adiabatic perturbation are $\log [10^{10} A_s^{\rm
adi}]=3.165\pm0.044$ and $n_s^{\rm adi}=0.972\pm0.014$ at 68\%
confidence level. The adiabatic spectrum index becomes larger, when
comparing with the constraint $n_s^{\rm
adi}=0.964\pm0.010~(1\,\sigma)$ from the pure adiabatic initial
condition. For the CDM isocurvature mode, the constraints are $\log
[10^{10} A_s^{\rm iso}]=-1.375^{+1.233}_{-1.306}$ and $n_s^{\rm
iso}=2.246^{+0.494}_{-0.428}$ (68\% C.L.). Furthermore, the
coefficient of the correlation is limited as
$-0.113<\cos\Delta<0.473$ at $95\%$ C.L.. Comparing with the minimal
$\chi^2$ of these two cases, we find that a weakly
positive-correlated mixture of CDM isocurvature and adiabatic
perturbations is mildly favored by the current data. However, the
fraction of CDM isocurvature mode can not be larger than $14.9\%$ at
95\% confidence level. The constraining power mainly comes from
WMAP7 temperature and polarization power spectra, since too
large fraction of the CDM isocurvature mode will lead to the
overstated modification of the shape of acoustic peaks in TT and TE
power spectra, as illustrated in Fig.\ref{inprinting_cmb_pk}. Our
results are consistent with the previous results reported in
Ref.\cite{Valiviita:2009bp}.

\begin{figure}[htbp]
\includegraphics[width=0.65\linewidth]{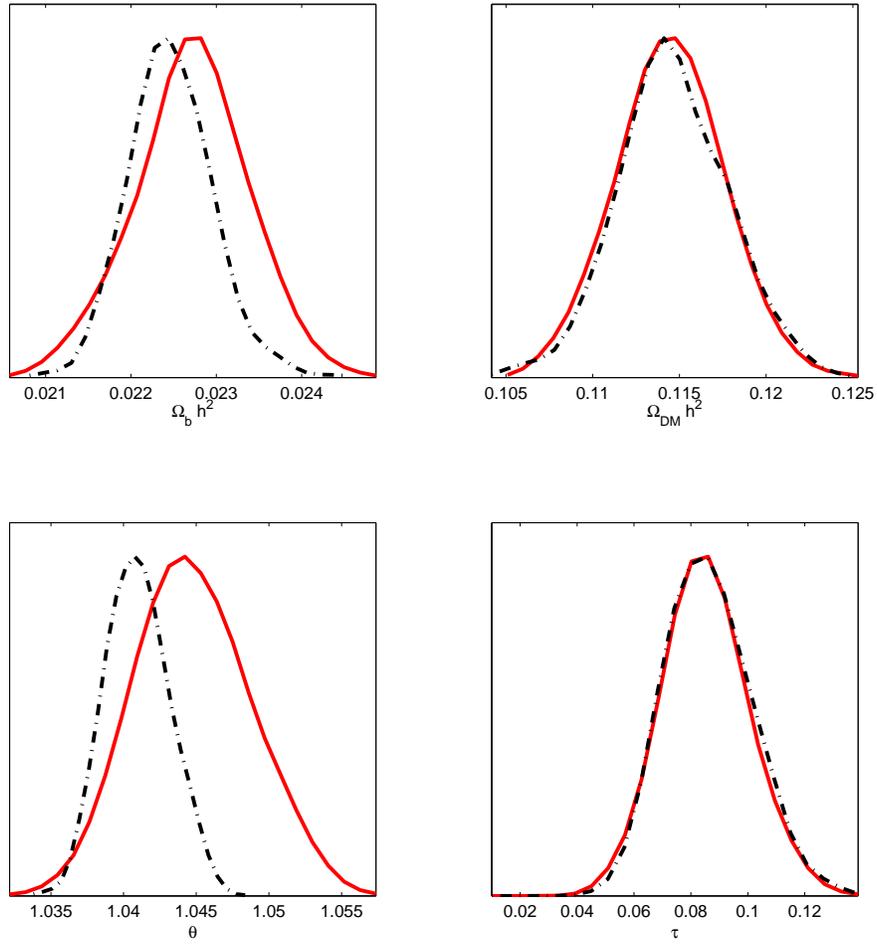}
\caption{One dimensional probability distributions of some
background cosmological parameters, the red solid lines are given by
the mixed initial condition, while the black dash-dot lines are from
the pure adiabatic conditions.}\label{fig_bkg}
\end{figure}

In Fig.\ref{fig_isopar}, we plot one dimensional posterior
probability distributions of $A^{\rm adi}_{s}$, $A^{\rm iso}_{s}$,
$n^{\rm adi}_{s}$ and $n^{\rm iso}_{s}$, as well as $\cos\Delta$ and $\alpha$ at the two different pivot scales. In Fig.\ref{fig_bkg} we also plot the one dimensional probability
distributions of the other cosmological parameters obtained from the
mixed initial conditions and pure adiabatic condition cases. The red
solid lines are given by the mixed initial conditions, while the
black dash lines are from the pure adiabatic one. One can see that
with the mixed initial conditions, the results favor a slightly
smaller $\omega_m$ and larger $\theta$ due to the correlation of the
CDM isocurvature perturbations with other cosmological parameters
when considering the non-adiabatic initial condition in our
analysis.

\begin{figure}[htbp]
\centering
\includegraphics[width=0.9\linewidth]{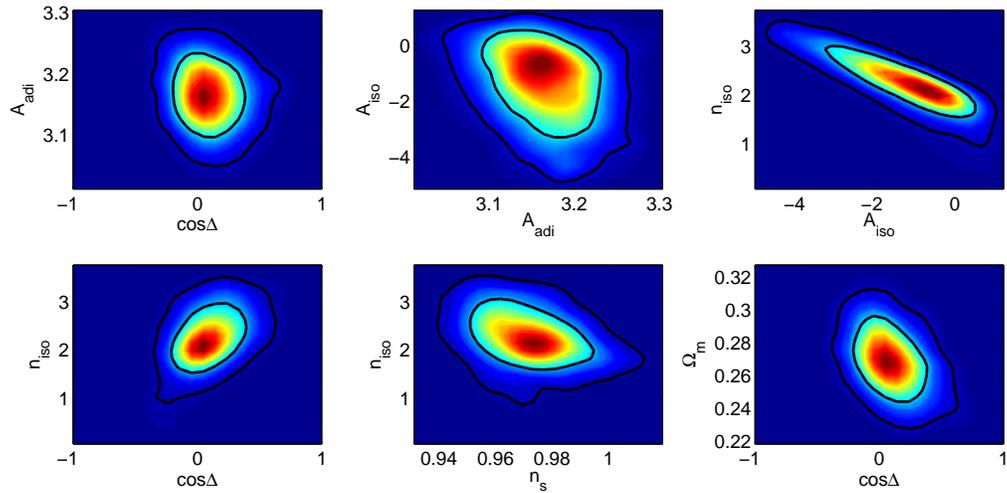}
\caption{Two dimensional contours among some cosmological
parameters.}\label{fig:dmiso}
\end{figure}

We find that there exist apparent correlations among the parameters related to initial conditions and other background cosmological parameters, as illustrated in Fig. \ref{fig:dmiso}. The coefficient $\cos\Delta$ is correlated with $\Omega_m$, since a positive correlated adiabatic and isocurvature component raise the peaks of TT spectrum which can be compensated by lower $\Omega_m$. A strongly negative correlation between $n_s^{\rm iso}$ and $A_s^{\rm iso}$ is shown in the right upper panel of Fig. \ref{fig:dmiso}.

\subsection{Information on reduced distance parameters}
\label{results_dp}

\begin{widetext}
\begin{table}[htbp]
\centering\caption{Median $1\,\sigma$ constraints on WMAP
normalization priors using full WMAP7 data for different
initial perturbations.}
\begin{tabular}{c|c|c|c|c|c}
\hline
\hline
 \multicolumn{2}{c|}{}& $R$ & $l_A$ & $r_s$ & $z_\ast$ \\
 \hline

 \multirow{2}{*}{WMAP only}&{Adiabatic}&  $1.721^{+0.018}_{-0.019}$ & $302.3\pm0.7$ & $146.7\pm1.5$ & $1091.1\pm0.9$ \\
\cline{2-6}
&{Mixed} &  $1.701_{-0.028}^{+0.027}$ & $301.5_{-1.6}^{+1.7}$ & $147.3\pm1.9$ & $1089.8\pm1.4$ \\
\hline

\multirow{2}{*}{Full data}&{Adiabatic} & $1.729\pm0.010$ & $301.8\pm0.6$ & $145.7\pm0.9$ & $1091.3\pm0.6$ \\
\cline{2-6}
& {Mixed} & $1.700_{-0.025}^{+0.026}$ & $300.1\pm1.3$ & $146.4\pm1.5$ & $1090.4\pm1.0$\\

\hline\hline
\end{tabular}\label{table:dp}
\end{table}
\end{widetext}

The WMAP normalization priors given by WMAP group include the
``shift parameter" $R$, the ``acoustic scale" $l_{A}$ and the photon
decoupling epoch $z_{\ast}$. $R$ and $l_A$ correspond to the ratio
of angular diameter distance to the decoupling era over Hubble
horizon and sound horizon at decoupling respectively, given by
\begin{eqnarray}
R(z_{\ast})&=&\sqrt{{\Omega_m}H_0^2}r(z_{\ast})~,\\
l_{A}(z_{\ast})&=&\pi\chi(z_{\ast})/r_{s}(z_{\ast})~,
\end{eqnarray}
where $r(z_{\ast})$ and $r_s(z_{\ast})$ denote the comoving distance
to $z_{\ast}$ and comoving sound horizon at $z_{\ast}$ respectively.
The decoupling epoch $z_{\ast}$ is given by \cite{Hu}
\begin{equation}
\label{defzstar} z_{\ast}=1048[1+0.00124(\Omega_b
h^2)^{-0.738}][1+g_1(\Omega_m h^2)^{g_2}]~,
\end{equation}
where \begin{equation} g_1=\frac{0.0783(\Omega_b
h^2)^{-0.238}}{1+39.5(\Omega_b h^2)^{0.763}},~
g_2=\frac{0.560}{1+21.1(\Omega_b h^2)^{1.81}}.
\end{equation}

\begin{figure}[htbp]
\includegraphics[width=0.65\linewidth]{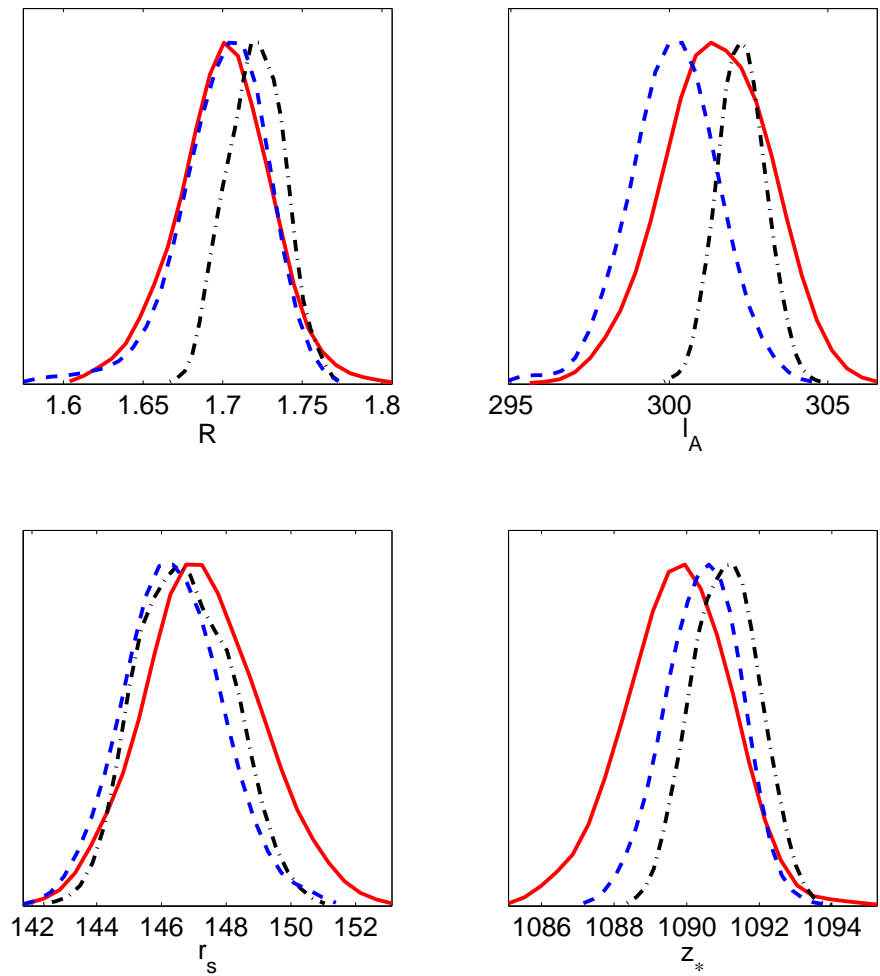}
\centering\caption{One dimensional posterior distributions of $l_A$,
$R$ and $z_{\ast}$ with WMAP7 data, the red solid lines are
given by the mixed initial perturbations, while the black dash lines
are from pure adiabatic perturbations.}\label{figuredp}
\end{figure}

The WMAP normalization priors encode in part of the CMB information
and can be used to constrain cosmological parameters to some extent.
They are derived parameters from the CMB power spectra based on the
fiducial cosmological model. Thus, they are model dependent
\cite{Li:2008cj,Wang:2007mza}. By using the WMAP7 data only,
in Fig. \ref{figuredp} we show the one-dimensional probability
distributions of the WAMP normalization priors in two different
initial perturbation conditions. The red solid lines are given by
pure adiabatic initial perturbations, while the black dash-dot lines
are with the mixed case. We find the obvious difference in the
probability distributions of $R$, $l_A$, and $z_{\ast}$. As we know,
the WMAP normalization priors mainly include the information on the
oscillatory structures of the CMB power spectrum. As shown in
Fig.\ref{inprinting_cmb_pk}, the TT power spectrum can be modified
by the CDM isocurvature perturbations, namely the locations and the
height of the peaks are modified when taking into account the
contributions of CDM isocurvature perturbations. And these effects
can lead to the difference of the mean values and errors of WMAP
normalization priors.

In table \ref{table:dp}, we list $1\sigma$ constraints on WMAP
normalization priors from the full WMAP7 data for different initial
conditions. One can see that, by taking into account the
contribution from the initial CDM isocurvature modes, constraints of
the WMAP normalization priors are obviously changed. The mean values
of $R$ is changed about $1\sigma$ when comparing with the pure
adiabatic case, while for $l_A$ and $z_{\ast}$, the changes are
about $1\sigma$ and $1.4\sigma$, respectively.

Furthermore, we also provide the $1\,\sigma$ constraints on $R$,
$l_A$, $z_{\ast}$ and $r_s$ using the combined data sample of WMAP7,
ACBAR, CBI, Boomerang, as well as SDSS LRG and SN Ia in table
\ref{table:dp}. As we expect, combining the data of SN and LSS can
help in constraining the other background cosmological parameters
and shrink error bars of $R$, $l_A$, $z_{\ast}$ and $r_s$
significantly. When considering the CDM isocurvature mode
perturbation, we find the differences between the constraints from
pure adiabatic perturbation and mixture of adiabatic and
isocurvature perturbations are improved obviously. In this case,
the difference of the mean values of $R$, $l_A$, $z_{\ast}$ and
$r_s$ become $2.9\sigma$, $2.8\sigma$, $1.5\sigma$ and $0.7\sigma$,
respectively.

\section{Summary}
\label{summary}

In this paper, we study the constraints on mixed adiabatic and isocurvature modes of cosmological perturbations. Using the current observational data, such as WMAP7 CMB power spectrum, matter power spectrum of SDSS data released seven LRG data and SNIa ``union2" sample, we find an adiabatic initial condition with the presence of certain isocurvature modes can explain the experiments better than a pure adiabatic initial condition. Moreover, we obtained more stringent constraints on the parameters of cosmological perturbations by virtue of the improvement of accuracy of observational data in recent years. Our result shows that the spectral index of isocurvature perturbation has a very blue tilt. This quantity may lead to either enhancement or depression of power spectrum on small scales, which depends on the correlation angle of isocurvature and adiabatic modes.

Given the WMAP normalization priors are widely used in performing testing of cosmological models, we provide the comparison on constraints of $R$, $l_A$ and $z_{\ast}$ from different initial conditions. Since WMAP normalization priors are derived parameters from the CMB power spectra, the constraints are changed obviously when the isocurvature mode modifies of the shape of the peaks and troughs in CMB power spectra.

As an end, we would like to mention that, from the point of view of information criteria, detailed constraints on cosmological parameters depend on the set of parameters of the model which is compared with observational data\cite{Liddle:2004nh}. The simplest model with purely adiabatic scale-invariant primordial power spectrum is able to capture the most relevant clues of early universe physics, however, more information deserves to be explored along with the improvement of observational data from forthcoming experiments, we expect the global analysis on both CMB power spectrum and matter power spectrum on small scales will be crucial to constrain the isocurvature perturbation.

\section*{Acknowledgements}

It is a pleasure to thank Andrew R. Liddle for comments. The work of HL and JL is supported in part by the National Natural Science Foundation of China under Grants Nos.11033005 and 10803001, and by the 973 program No. 2010CB833000 and by the Youth Foundation of the Institute of High Energy Physics under Grant No. H95461N. YFC thanks the Department of Physics of Mcgill University for the hospitality when this work was finalized. The research of YFC is supported in part by the Arizona State University Cosmology Initiative.

\end{document}